\documentclass[12pt,amssymb]{article}
\newcommand{\be}{\begin{equation}}
\newcommand{\ee}{\end{equation}}
\newcommand{\bea}{\begin{eqnarray}}
\newcommand{\eea}{\end{eqnarray}}

\begin{document}

\begin{titlepage}
\begin{center}
\vskip .2in
\hfill
\vbox{
    \halign{#\hfil         \cr
           hep-th/0101010 \cr
           SU-ITP-0038 \cr
           UCSD-PTH-00-35\cr
           December 2000    \cr
           }  %end of \halign
      }   %end of \vbox
\vskip 1.5cm
{\Large \bf Comments on Unstable Branes}\\
\vskip .3in
{\bf Ken Intriligator,}
\footnote{email address:keni@ucsd.edu}
{\bf Matt Kleban}
\footnote{e-mail address:mkleban@itp.stanford.edu}
and {\bf  Jason Kumar}
\footnote{e-mail address:j1kumar@ucsd.edu}\\
\vskip .25in
{\em
${}^{1,3}$ Department of Physics,
University of California, San Diego\\
La Jolla, CA  92093-0354 USA \\}
\vskip .1in
{\em
${}^2$ Department of Physics,
Stanford University\\\
Stanford, CA  94305 USA\\}
\vskip 1cm
\end{center}
\begin{abstract}

We argue that type II string theories contain unstable NS4 branes,
which descend from a conjectured unstable M4 brane of M-theory.
Assuming that an M2 brane can arise in M5 brane/anti-brane
annihilation, the unstable M4 brane, and also an unstable M3 brane,
must exist as sphalerons.  We compare the tensions 
of the unstable NS4 branes, M4
brane, and related type II unstable D-branes, and present 11d
supergravity solutions for unstable Mp branes for all $p$.
We study the $Z_2$ gauge symmetry on the
worldvolume of unstable branes, and argue that it can never be
unbroken in the presence of lower brane charge.

\vskip 0.5cm
 \end{abstract}
\end{titlepage}
\newpage

\section{Introduction}

The study of unstable brane systems has been an important direction in
string theory research, following the work of \cite{Sen1,Senrev}.  In the
context of type IIA(B) string theory, simple examples are a
Dp-$\overline{\rm Dp}$ system and a single D($p-1$) brane, with $p$
even (odd).  These systems can decay to the closed string vacuum or to
a state with some stable branes of lower dimension, as reviewed in the
next section.

In this work we consider aspects of unstable brane systems and the
$SL(2,Z)$ duality of type IIB from the point of view of M theory.
For the most part, the study of M-theory has focused on the study
of stable supersymmetric objects.  However, there are compelling
theoretical and phenomenological reasons for attempting to
understand configurations in which supersymmetry is broken.
Historically, the understanding of the structure of the
atom depended greatly on the study of radioactive decay
processes.  It would be interesting to see if,
in a similar way, the study of the decay of unstable branes
gives any insight into the nature of the
fundamental degrees of freedom of M-theory.

One might worry that it would be hopeless to follow non--BPS
configurations from small to large string coupling.  However, as discussed
in \cite{Sphal}, unstable branes can be regarded as sphalerons whose
existence is protected by BPS charges.  Perhaps, then, it is sensible
to extrapolate to strong coupling.  Other works which consider
unstable branes in the context of S-duality and M-theory include
\cite{LHYL}, which considered general unstable Mp branes.

In IIB string theory we can S-dualize the two sides of
Dp-$\overline{\rm Dp}\rightarrow$ D($p-2$).  We obtain, for
example, NS5-$\overline{\rm NS5}\rightarrow$ D3 (more generally, we
can consider $(p,q)$ 5-branes).  Likewise, we can S-dualize the
unstable D(2n)-branes to unstable NS(2n)-branes or, more generally,
unstable $(p,q)$ 2n-branes. We'll focus on the unstable NS4 brane.
The unstable NS4 can decay to the vacuum or, if the
tachyon field condenses in a kink configuration, to a D3 brane.

By compactifying the IIB theory and T-dualizing to IIA, we argue that
the unstable NS4 of IIB implies the existence of
 an unstable NS4 brane of IIA.  The
unstable IIA NS4 brane can decay to the vacuum or, if there are
vortices, to D2 branes.  In this way, the single unstable IIA NS4 is
similar to the unstable D4-$\overline{\rm D4}$ system, though the
vortex is associated with the center of mass $U(1)$.  This suggests
that there can be an intermediate unstable NS3, in analogy with the
IIA unstable D3, which can also decay to the D2 brane via a kink.

We also consider uplifting unstable IIA brane systems to M-theory.
The unstable IIA NS4 brane should correspond to an unstable M4
brane.  The unstable M4 can decay to the vacuum or, if there are vortices
in the worldvolume gauge field,
to M2 branes.  As with the NS3 brane mentioned above, we should also
expect an intermediate unstable M3, which can decay to a M2 brane via
a kink.  As discussed in \cite{Piljin}, the M2 can arise from
M5-$\overline{\rm M5}$ annihilation with a unit of $H_3$ flux.  The
unstable M4 and M3 can then be regarded as sphalerons, along the lines
of \cite{Sphal}, whose existence is protected by the same non-trivial
configuration space topology as leads to the M2 in M5-$\overline{\rm
M5}$ annihilation.  

The conjectured unstable NS4 branes of IIA and IIB string theory are
discussed in sect.3.  The unstable M4 and M3 branes are
discussed in sect. 4.  In sect. 5 the connection between the unstable
M-branes and IIA branes is discussed in more detail.  The unstable
IIA D3 brane, for example, is interpreted as an unstable M4 brane
wrapped along the dilaton direction.  The tensions of these unstable
branes are compared and the action of IIB S-duality is further
discussed.  Section 6 contains 11d SUGRA solutions associated with
the unstable M-branes, along the lines of \cite{BMO}.

Section 7 is devoted to a somewhat separate topic: the $Z_2$ symmetry
of unstable type II branes, under which the real tachyon transforms as
$T\rightarrow -T$.  This $Z_2$ is actually a gauge symmetry (a fact
which played an important role in \cite{HKL}), which is Higgsed in the
vacuum by the nonzero $T$ expectation value.  We discuss some issues
associated with the $Z_2$ gauge symmetry, such as the fact that the
worldvolume coupling $\int dT\wedge C$ would not be $Z_2$ gauge
invariant unless the RR field $C$ is also $Z_2$ charged, $C\rightarrow
-C$ (this was also suggested to us by A. Sen \cite{Senc}).  A nonzero
$C$ expectation value then also spontaneously breaks the $Z_2$ gauge
symmetry.

\section{Review}
As is well known, a coincident Dp-$\overline{\rm Dp}$ brane system has
a worldvolume $U(1)\times U(1)$ gauge theory, which contains a complex
scalar field $T$ charged under the relative $U(1)$.  The worldvolume
theory is an Abelian Higgs model, with a ``Mexican hat'' potential for
$T$.  The potential has a local maximum at $T=0$ and a global minimum
at $|T|=T_0$, with $V(|T|=T_0)+2T_p=0$, so that the energy of the
condensed ``tachyon'' $T$ cancels the tension of the Dp-$\overline{\rm
Dp}$ system.  Having $T$ relax to some fixed $\langle T\rangle $, with
$|\langle T \rangle |=T_0$, is interpreted as annihilation of the
brane-antibrane system to the closed string vacuum
\cite{Senrev}.

More generally, a Dp-$\overline{\rm Dp}$ pair, with $n$ units of
magnetic flux, carries the charge of $n$ D($p-2$) branes, which is the
final product of the brane annihilation \cite{Sen1}.  Upon adding the
$n$ units of magnetic flux to the worldvolume Abelian Higgs theory,
the finite energy solution is a codimension 2 vortex, with $T$ winding
precisely $n$ times around its minimum:
 \be n={1\over 2\pi}\oint A={1\over 2\pi i}\oint
d(\ln T), \ee where the integral is over a large circle surrounding the
vortex.  This correlation between the number of units of magnetic
flux and the winding number of $T$ follows from the fact that the
vortex configuration only has finite energy if the covariant
derivative of $T$ vanishes far {}from the vortex, $(\partial _\mu
+iA_\mu)T\rightarrow 0$ as $r\rightarrow \infty$.  So if, for example,
there is $n=1$ unit of D($p-2$) brane charge, there is a
vortex configuration where $T$ winds precisely once; likewise, if
there is no D($p-2$) brane charge, $n=0$, there cannot be a finite
energy vortex configuration with winding $T$:  only the trivial vacuum
is allowed.  The vortex carries the Ramond-Ramond charge of $n$ D($p-2$)
branes thanks to a worldvolume coupling $\sim \int F\wedge C$.

Sen further argues that one can consider an intermediate unstable
D($p-1$) brane in the decay of Dp-$\overline{\rm Dp}$ to the vacuum or
a D($p-2$) brane.  The intermediate unstable D($p-1$) brane has a real
tachyon $T$ and an even potential $V(T)=V(-T)$, with $T=0$ unstable,
and stable minimum at some $|T|=T_0\neq 0$.  The vacuum where $|T|=
T_0$ is interpreted as the closed string vacuum, with $V(T_0)$
canceling off the D$(p-1)$ brane tension: $V(T_0)+T_{p-1}=0$.  The
tachyon can also condense to a kink configuration, with $T\rightarrow
+T_0$ for $x_t\rightarrow +\infty$ and $T\rightarrow -T_0$ for
$x_t\rightarrow -\infty$ (where $x_t$ is the transverse coordinate to
the codimension 1 domain wall). The domain wall is interpreted as the
D$(p-2)$ brane to which the unstable D($p-1$) brane can decay.  In the
type II context, there must be a world-volume coupling
\cite{Senrev}\be \int dT\wedge C_{p-2}, \ee in order for the $T$ kink
configuration to carry the Ramond-Ramond charge of a D$(p-2)$ brane.

\section{Unstable NS-4-Branes}

We start by considering the S-dual of the D5-$\overline{\rm
D5}\rightarrow {\rm D3}$ brane decay process in Type IIB string
theory.  Although it is notoriously difficult to follow
non-supersymmetric states through S-duality, one nevertheless can easily
understand both the beginning and end of the condensation path.
  By widely separating the D5-$\overline{\rm D5}$, the system
becomes meta-stable and S-dualizes to a NS5-$\overline{\rm NS5}$ brane
system.  The S-dual of the final product D3 brane is again a
D3 brane.  So a coincident NS5-$\overline{\rm NS5}$-brane pair,
with magnetic flux turned on, will decay to D3-branes.  In analogy
with the unstable D4, we claim that there exists an unstable NS4, which
can be regarded as an intermediate state in the NS5-$\overline{\rm
NS5}\rightarrow D3$ decay process.

The IIB NS5-$\overline{\rm NS5}$ worldvolume theory is a $U(1)\times
U(1)$ gauge theory, with complex tachyon charged under the relative
$U(1)$, in complete analogy with the D5-$\overline{\rm D5}$.  There
should be an unstable local maximum of the potential at $T=0$ and
global minimum at $|T|=T_0$, with $2T_{NS5}+V(T_0)=0$,
canceling off the tension of the two NS5 branes.  The branes can
annihilate to the vacuum or, with $n$ units of magnetic flux
in the relative $U(1)$, there is a codimension 2 vortex, with
winding number $n$, which is interpreted as the $n$ D3 branes
to which the NS5-$\overline{\rm NS5}$ (with $n$ units of flux)
decays.

In the NS5-$\overline{\rm NS5}\rightarrow$ D3 vortex configuration,
the tachyon may happen to condense into two domains separated by an
unstable domain wall.  If the tachyon condenses to positive real
values in one domain, and negative real values in the other, then it
interpolates through $T=0$ in the domain wall.  $T=0$ corresponds to
the state where the tachyon has not condensed and is described by the
NS5-$\overline{\rm NS5}$, where D-strings can end.  So, in analogy
with the heuristic arguments for the unstable D4-brane \cite{Senrev},
IIB string theory has an unstable 4+1 object, on which D-strings can
end; this is the IIB unstable NS4 brane.  The NS5-$\overline{\rm NS5}$
brane system can decay to an unstable NS4-brane by the condensation of
one real tachyon and the NS4, in turn, decays to a D3-brane by the
kink condensation of another real tachyon.

We can now consider the T-dual of this system.  Suppose that one
direction, which is common to each brane in NS5-$\overline{\rm
NS5}\rightarrow\rm{NS4}\rightarrow\rm{D3}$, is taken to be compact
(there cannot be a vortex on a compact direction without an
anti-vortex).  T-dualizing along this circle gives a IIA system, with
a coincident NS5-$\overline{\rm NS4}\rightarrow
\rm{NS4}\rightarrow\rm{D2}$.  The fact that the unstable IIB NS4 brane
gets T-dualized to a IIA unstable NS4 brane can be seen by studying
the D-strings which begin and end on it.  Under T-duality along a
longitudinal direction, they become D2-branes which begin and end on a
4+1 dimensional hypersurface.  In analogy with the IIA NS5-brane (on
which membranes can also begin and end), we call this surface the IIA
unstable NS4-brane.  Membranes intersects the NS4 brane on a
world-volume string, which couples to a world-volume two-form tensor
gauge field.  In 4+1 dimensions, strings coupled to two-form gauge
fields are electric-magnetically dual to particles coupled to ordinary
gauge fields.  We therefore conjecture that both the IIA and IIB NS4
branes have a world-volume $U(1)_c\times Z_2$ gauge theory, where the
$U(1)_c$ is the ``center of mass'' $U(1)$ of the NS5-$\overline{\rm
NS5}$ system.  In the IIB case, the stable D3 brane comes from a $Z_2$
kink.  In the IIA case, the stable D2 brane comes from a vortex in
$U(1)_c$, which Higgses $U(1)_c$.  The $U(1)_c$ charged tachyon here
is a nonperturbative state, along the lines of
\cite{Piljin,confine}.

\section{Unstable M4 and M3-Branes}

We now consider lifting the type IIA decay NS5-$\overline{\rm
NS5}\rightarrow \hbox{NS4}\rightarrow D2$ to M-theory.  The
NS5-$\overline{\rm NS5}$ becomes a M5-$\overline{\rm M5}$, at a point
on the dilaton circle.  Likewise, the D2 becomes an M2 at a point on
the dilaton circle.  We therefore expect that the intermediate
unstable NS4 becomes an unstable M4 brane, again at the same point on
the dilaton circle.  We thus have an M5-$\overline{\rm M5}$-brane
pair, with an appropriate $H_3$ magnetic flux, decaying to an unstable
M4 brane, which will further decay to an M2-brane.  The direct decay
of the M5-$\overline{\rm M5}$-brane pair, with $H_3$ flux, to a
M2-brane was described in
\cite{Piljin}.

The M5-$\overline{\rm M5}$-brane pair has a $U(1)_1\times U(1)_2$
tensor multiplet theory living on its worldvolume, with tachyonic
strings (which are charged under the relative $U(1)$) coming {}from M2
membranes stretched between the 5-branes.  The M4 brane is then also a
place where M2 membranes can end, so its worldvolume gauge theory
naturally has two-form gauge fields, with charged strings.  The
expected gauge group is $U(1)_c\times Z_2$, which is analogous to that
of type II unstable branes.  As with the IIA NS4 brane, in 5d we can
Hodge dualize the two-form gauge fields and electrically charged
strings to ordinary gauge fields coupled to electrically charged
particles.  Thus the $U(1)_c$ in the unstable M4 worldvolume can be
described as an ordinary gauge theory, with electrically charged
particles.  (The worldvolume $Z_2$ gauge symmetry is perhaps more
complicated, since the Hodge-duality between tensors and vectors need
not directly apply to discrete gauge symmetries, which do not have
propagating gauge fields.)

We expect that the M4 worldvolume gauge theory has a complex tachyon,
which is electrically charged under the $U(1)_c$ gauge group. This is
similar to the $U(1)_c$ charged non-perturbative tachyon of
\cite{Piljin,confine}, whose condensation was argued to confine
$U(1)_c$.  In the present context, a configuration with $U(1)_c$
magnetic flux can have a codimension two vortex, where this complex
tachyon winds around its $S^1$ vacuum configuration space.  This
vortex is the M2 brane to which the M4 brane can decay.

As usual, we might expect that the M4$\rightarrow$ M2 decay, via a
vortex, can occur via an intermediate unstable state: an unstable M3
brane, which can decay to the M2 brane via a kink in its worldvolume
real tachyon scalar. Since there is no stable M1 brane, the M3 brane
worldvolume gauge theory must not admit stable codimension 2 vortices.

The M4 and M3 branes can be regarded as sphalerons\footnote{See
the recent discussion in \cite{Sphal} (in particular Figs. 3 and 4),
and the references cited therein, for the relevant background on
sphalerons.}, whose existence is related to that stable M2 branes.
Consider the scenario of \cite{Piljin}, where a M5-$\overline{\rm M5}$
with $H_3$ flux can decay to a stable M2 brane.  The M2 is some
codimension 3 solution for the fields in the M5-$\overline{\rm M5}$
worldvolume; its stability should be protected by non-trivial $\pi _2$
topology.  We can now write the same solution, but with some of the
three transverse coordinates to the M2 replaced with parameters.
Replacing one coordinate with a parameter $\sigma$, which lives on a
circle $\sigma \in [0,2\pi]$, we get a 1-parameter family of
codimension 2 solutions.  Taking the vacuum to be at $\sigma =0$, the
M3 brane is the sphaleron at $\sigma =\pi$.  The one negative mode,
associated with sliding down the circle, is interpreted as the real
tachyon of the M3 brane, which can lead to a M2 brane via a kink.

Similarly, we can replace two of the coordinates in the M2 brane
solution with parameters living on an $S^2$, to obtain a family of
codimension 1 configurations.  Taking the configuration at the south
pole of the $S^2$ to be the vacuum, the configuration at the north
pole is our unstable M4 brane sphaleron.  There are two real unstable
modes, associated with sliding down the $S^2$, which are interpreted
as the complex tachyon in the M4 brane worldvolume; a vortex in this
tachyon leads to the M2 brane.  The existence of the unstable M3 and
M4 branes are thus guaranteed by the same non-trivial $\pi _2$ topology
which ensures the stability of the BPS M2 brane configuration in the
M5-$\overline{\rm M5}$ annihilation process.

The existence of the M4 brane is also suggested\footnote{We thank
O. Bergman for pointing this out to us.}
by the work of \cite{OBMG} on non-BPS states in the heterotic theory
on $T^4$ and type IIA on K3.  The dual of the $Z_2$ valued D-particle
seen in the type I string was argued there to arise from a non-BPS
4-brane in IIA or M theory, which is wrapped on the $T^4/Z_2$.

In \cite{MFPH} it was further suggested that end-of-the-world
M9-$\overline{\rm M9}$ branes can annihilate to form M5 branes,
which are associated with $\pi _3(E_8)=Z$.  If so, 
along the lines of the discussion above, one should also
expect unstable M8, M7, and M6 branes as sphalerons, whose existence
is protected by the same $\pi _8(E_8)$ topology as gives the M5 brane.

\section{Type II Branes from Unstable M-branes}

Consider the decay M5-$\overline{\rm M5}\rightarrow \rm{M4}
\rightarrow \rm{M2}$, but now with one direction common to each of the
branes wrapped along the dilaton direction.  The wrapped M5-brane is
the IIA D4-brane, the wrapped M2 brane is the IIA fundamental string,
and we identify the wrapped M4 brane as the IIA unstable D3-brane.  So the
IIA decay is D4-$\overline{\rm D4} \rightarrow {\rm D3}\rightarrow
\rm{F1}$.  The IIA unstable D3 has a worldvolume $U(1)_c\times Z_2$
gauge theory, and the fundamental string arises via a $U(1)_c$ vortex.
The $U(1)_c$ charged tachyon which winds in this vortex must arise
non-perturbatively, along the lines of \cite{Piljin,confine}.  The
corresponding M4$\rightarrow$ M2 decay presumably also occurs via a
vortex, with winding charged tachyon vev, in the dualized $U(1)_c$ 
M4 worldvolume gauge theory.

A more standard IIA decay is D4-$\overline{\rm D4}\rightarrow
\rm{D3}\rightarrow D2$, with the D4-$\overline{\rm D4}$ system having
a unit of magnetic flux in the relative $U(1)_r$ and the $Z_2$
charged, perturbative, real tachyon in the D3 brane worldvolume (which
is $U(1)_c$ neutral) having a kink.  This decay also corresponds to
the M-theory decay M5-$\overline{\rm M5} \rightarrow
\rm{M4}\rightarrow \rm{M2}$, with the M5-$\overline{\rm M5}$ and
the M4 again wrapped on the dilaton circle, but now with the M2 brane
at a point in the dilaton direction.  Since D3$\rightarrow $D2 occurs
via a kink in the D3 worldvolume $Z_2$ gauge theory, presumably its
M-theory uplift to M4$\rightarrow $M2 should also involve a kink in
the M4 worldvolume $Z_2$ tensor (rather than vector, and thus perhaps
mysterious) gauge theory.  This kink is presumably what distinguishes
this case {}from the above M4$\rightarrow $M2 decay, corresponding to
D3$\rightarrow $F1.

We now consider the tensions of the unstable branes.  At weak
coupling, the tension of a stable type II Dp brane is
$T_p=1/g(2\pi)^pl_s^{p+1}$, whereas the tension of an unstable type II
Dp brane is $T_p=\sqrt{2} /g(2\pi)^p l_s^{p+1}$ \cite{Senrev};
numerical factors will henceforth be dropped. IIA tensions are lifted
to M-theory using $g\propto R/l_s$ and $gl_s^3 = l_{11}^3$.

Interpreting the IIA unstable D3 as an unstable M4 wrapped on the
dilaton circle, one would naively expect its tension to be given by
$T_{D3}=\int_0 ^{2\pi R} T_{M4}$, with $R$ the radius of the dilaton
direction and $T_{M4}$ the tension of the unwrapped, unstable
$M4$-brane.  Using $T_{D3}\propto 1/gl_s^4$ for $g\ll 1$, this gives
\be T_{M4}\propto 
{1\over l_{11}^{9\over 2} R^{1\over 2}}\qquad\hbox{for}\quad R\ll l_{11},
\ee
where the $R \ll l_{11}$ restriction is because the D3 tension is only
known for $g\ll 1$.  The fact that the above tension is $R$ dependent
might seem troubling, but is perhaps not entirely surprising. If, for
example, the unstable M4 were placed on a transverse circle, one might
expect an interaction energy between the M4 and its images.  Of course
this would not happen with BPS states, but for the unstable branes
there is no supersymmetry to impose a no-force condition.  So the
effective tension of the M4 could depend on the radius of the
transverse circle.  Similarly the tension of the wrapped M4 could be
expected to be radius dependent, as given above.

Naively applying the tension formula (3) outside of its region of validity
would suggest that $T_{M4}\rightarrow 0$ for $R\rightarrow \infty$.
Assuming that the M4 exists as an unstable state in uncompactified M-theory,
we would expect that (3) is not exact for all
$R$, with $T_{M4}\sim 1/l_{11}^5$ in the $R\rightarrow \infty$ limit.

Consider wrapping the IIA unstable D3 brane on a circle of radius
$R_2$.  In M-theory we expect this to correspond to the M4 brane
wrapped on a torus, with sides of length $R_1\sim g l_s$ (the dilaton
direction) and $R_2$.  For $g\ll 1$, we know that the tension of the
unwrapped D3 is $T_{D3}\propto 1/gl_s^4$.  Wrapping the D3 on the circle
of radius $R_2\gg l_s$ gives a 2+1 dimensional object with tension
\footnote{By $T_{p+1}$, we shall mean the tension of the p+1 dimensional
object obtained by considering the brane on a length scale such that
all other (compact) directions are small.  This tension has dimensions
of $[energy]^{p+1}$.}
$T_{2+1}
\propto R_2T_{D3} \propto R_2/gl_s^4$ which, in terms of M-theory,
is \be T_{2+1}\propto {R_2\over gl_s^4} ={R_2 R_1^{1\over 2} \over
l_{11}^{9\over 2}}\qquad\hbox{for}\quad R_1 \ll l_{11} \ll R_2.  \ee
The expected invariance under exchanging $R_1$ and $R_2$ must be
restored by corrections to (4) which are subleading in the above
$R_1\ll l_{11} \ll R_2$ limit.  Upon T-dualizing, this wrapped IIA D3
brane becomes an unwrapped, unstable IIB D2 brane.  The requirement
of invariance under exchanging $R_1$ and $R_2$ is then S-duality in
the IIB coupling $g_b=R_1/R_2$.

We next consider the type II unstable NS4 branes.  A guess for their
tension at weak coupling is $T_{NS4}\propto 1/g^2 l_s^5$.  This is
compatible with the T-duality between the wrapped IIA NS4 and wrapped
IIB NS4, since $R_a/g_a^2l_s^{5}=R_b/g_b^2l_s^5$ (using
$g_b=g_al_s/R_a$ and $R_b=l_s^2/R_a$).

Assuming the $R_1\leftrightarrow R_2$ invariance, we can use (3) to
obtain the tension of the IIA NS4 wrapped on a small circle of radius
$R_2$, with $R_2\ll l_{11}\ll R_1$: \be T_{3+1_{NS4-IIA}} \propto
{R_2^{1\over 2} \over l_{11}^{9\over 2}} = {R_2^{1\over 2} \over
g_a^{3\over 2} l_s^{9\over 2}} \qquad \hbox{for} \quad R_2\ll
l_{11}\ll R_1.  \ee The tension (5) is a prediction for strong IIA
coupling, $g_a\sim (R_1/l_{11})^{3/2}\gg 1$; it differs from the above
weak coupling guess $R_2/g_a^2l_s^5$.  Now T-dualize, to get the IIB
NS4 brane, wrapped on a circle of radius $R_b=l_s^2/R_2$.  In terms of
$R_b$ and the IIB coupling $g_b=R_1/R_2$, the tension (5) of this
wrapped object is \be T_{3+1_{NS4-IIB}} \propto {R_b \over g_b
^{3\over 2} l_s^5} \qquad \hbox{for}\quad g_b\gg 1, \ \ R_b\
\rm{large}.  \ee Unlike the (strong coupling, small wrapping radius)
tension (5), the (strong coupling, large radius) tension (6) has
standard dependence on the wrapping radius $R_b$.  Thus
$T_{3+1_{NS4-IIB}}=\int _0^{2\pi R_b}T_{NS4-IIB}$ gives \be
T_{NS4-IIB} \propto {1\over g_b^{3\over 2} l_s^5} = {1\over
g_b^{1\over 4} l_{10p}^5} \qquad\hbox{for}\quad g_b\gg 1, \ \ R_b\
\rm{large}, \ee a radius independent tension.  $l_{10p}$ is the IIB,
$SL(2,Z)$ invariant, 10d Planck length.  The tension (7) could be
computed another way: its S-dual is \be T_s \propto {g_b^{1\over 4}
\over l_{10p}^5}= {1\over g_b l_s^5} \qquad\hbox{for}\quad g_b\ll 1,\
\ R_b\ \rm{large} \ee which is just the tension of the IIB unstable
D4-brane.  As might have been anticipated, the unstable $NS4$-brane of
IIB is the S-dual of the unstable D4-brane of IIB.

The reader might wonder if this compatibility of the $T_{D4}$ and
$T_{NS4}$ with S-duality really had any right to work.  The above
wrapped IIA NS4 brane tension (5) is only the leading order term in an
expansion in $R_2/l_{11}$ and $R_2/R_1$.  On the other hand, consider
starting with the IIB unstable D4, whose tension is only known to
leading order in $g_B=R_1/R_2$ and $R_b$ large.  Upon S-dualizing,
this gives the IIB NS4 brane tension to leading order in
$1/g_B=R_2/R_1$ and $R_b$ large.  Upon T-dualizing, this gives the IIA
NS4 brane tension in the limit of $R_2$ small compared to $R_1$ and
$l_{11}$, which agrees with the region where our analysis of the
wrapped IIA NS4 is valid.  The compared tension is just the leading
term of a double expansion, in $R_2/l_{11}$ and $R_2/R_1$, for both
the wrapped NS4 of IIA and NS4 of IIB.  The higher order corrections
presumably also match, so that the NS4 of IIA and IIB are T-dual, and
the NS4 and D4 of IIB are S-dual, for all coupling and radii.

\section{Unstable branes in 11d SUGRA}

In this section we find non-BPS solutions of 11d SUGRA corresponding
to uncharged Mp branes, for arbitrary p.  The ansatz for the Mp 
brane metric, in isotropic coordinates, is 
\be ds^2 = e^{2A(r)}\left( -dt^2 + dx_a dx^a \right) + e^{2B(r)}
\left( dr^2 + r^2 d \Omega_{9-p}^2 \right), \ee where $a$ runs
{}from 1 to $p$, with the 4-form $H_4=0$ (since there is no charge).
Einstein's equations then imply $R_{\mu\nu}=R=0$, which
amounts to a set of differential
equations for the functions
$A(r)$ and $B(r)$:
\be
A'' + (p+1) A'^2 - (p-8) A' B' - (p-9) A' / r = 0,
\ee
\be
(p-9) B'' - (p+1) \left[A'' + A' \left(A' -  B' \right)\right] + (p-9) B'/r =0,
\ee
\be
B'' + (8-p) B'^2 + (p+1) A' \left( B' + 1/r\right) + (17 -2p) B'/r=0.
\ee

The most general solution to these equations is
\be e^{2A(r)}=
\left({f_{-}(r) \over f_{+}(r)}\right)^C, \quad 
e^{2B(r)} = \left({f_{+}(r)
\over f_{-}(r)}\right)^{C(p+1)\over (8-p)}[
f_{+}(r)f_{-}(r) ]^{2/(8-p)},
\ee
where \be f_{\pm}(r) \equiv 1 \pm \left({r_0 \over r}\right)^{8-p},
\quad\hbox{}\quad C \equiv {2\over 3} \sqrt{9-p\over p+1}, \ee and
$1/r_0$ is the mass parameter.  We can convert from the above isotropic
coordinate $r$ to a Schwarzschild coordinate $\rho$ via $\rho =e^{B}r$.
The $p=0$ case of the above general
solution is then recognized as the 11d Schwarzschild metric 
\be
ds^2=-\left( 1-{4r_0^8\over \rho ^8}\right) dt^2+\left( 1-{4r_0^8\over \rho
^8}\right) ^{-1}d\rho ^2 +\rho ^2 d\Omega _9^2.  
\ee 

The above metric (9), (13) properly asymptotes to flat 11d spacetime in the
$r\rightarrow \infty$ limit (provided $p<8$).  The tension $T_p$ of
the Mp brane can be found by expanding in the $r\rightarrow \infty$
limit \be g_{tt}\rightarrow -1+const. {l_{11}^9T_p\over
r^{8-p}}+\dots; \ee this gives \be T_{p}\propto {r_0^{8-p}\over
l_{11}^9}.  \ee Note also that the above solution only makes
sense for $r\geq r_0$, and has a singularity at $r=r_0$.  For $p=0$
this is merely a coordinate singularity associated with the
Schwarzschild horizon.  For $p>0$, $r=r_0$ is truly singular, as can
be seen in the curvature invariant $R_{Riemann} = R^{\mu \nu \lambda
\sigma} R_{\mu \nu \lambda \sigma}.$

As might have been expected from Birkhoff and no-hair theorems, the
above Mp brane solutions contain only a single parameter, the mass
$1/r_0$.  To see this, note that the first of the three differential
equations can be used to solve for $B'$ in terms of $A'$ and $A''$.
Integrating $A'$ to $A$ and $B'$ to $B$ introduces trivial constants,
which are fixed by the boundary condition that the solution approach
the Minkowski solution at large $r$.  That leaves two second order
equations for a single variable.  The first can be solved, with two
undetermined constants.  The second differential equation fixes one of
these, leaving $r_0$ as the only free parameter.

The above 11d SUGRA solutions are analogous to type II SUGRA
unstable brane solutions found in \cite{BMO}.  However, the
solution of \cite{BMO} depended on a second parameter, in addition
to the mass $1/r_0$.  An argument similar to the one above
indicates that the 10D solution should indeed depend
on an additional parameter, due to the fact that the dilaton
field can vary independently of the metric.  No-hair theorems
fail in this case, due to the presence of a curvature singularity.
The additional parameter was interpreted in \cite{BMO} as
corresponding to the expectation value of the tachyon field.  

A possible reason for why the type II case differs from
M-theory (in having a parameter corresponding to the tachyon vev) is that
IIA unstable branes are long lived at weak coupling.  This is because
the IIA unstable branes are expected to have a decay time on order
$l_s$ which, at weak coupling, is long lived compared to the natural
SUGRA time scale $l_{10p}=g^{1/4}l_s$.  On the other hand, 
unstable Mp branes should be expected to decay on time scales
on order $l_{11}$, and thus meta-stable SUGRA solutions including the
extra tachyon parameter should not be expected.  

KK reducing the 11d SUGRA action leads to the IIA SUGRA action via \be
G = \left[ \begin{array}{cc} \lambda^{-1/8} g & 0 \\ 0 & \lambda
\end{array} \right],
\ee
where $G$ is the metric in 11 dimensions, $g$ is the
10 dimensional metric, and $\lambda$ the dilaton.  KK reducing the above
unstable Mp brane solutions then leads to unstable brane solutions
of IIA SUGRA, along the lines of \cite{BMO}.  The metric ansatz and
requirement that there be an isometry along the reducing
direction implies that one can only dimensionally reduce these solutions
along the brane, i.e. reducing the Mp solution to a D(p-1) solution.  
The solution of \cite{BMO} is then indeed obtained
\footnote{The Mp solution reduces to the D(p-1) solution of
\cite{BMO}, with their parameters set to $c_2 =-1$ and $c_1 =
3C{q-4\over 7-q}$, with $q=p-1$, and we need to choose the negative
branch in defining their parameter $k=-{3\over 4} Cq$. In the
Euclidean version, as in \cite{Sphal} (see their fig. (2)), the
D(p-1) brane has periodic fermion boundary conditions around
the Euclidean time $S^1$.}.  Of course, the classical 11d SUGRA solutions
should really only be trusted when the dilaton circle is very large.
They can thus be regarded as the unstable brane solutions of IIA
string theory at very strong string coupling.  On the other hand, the
solutions of \cite{BMO} should really only be trusted for weak string
coupling, when the dilaton circle is very small.  So we should not,
for example, expect the ADM tension $T_p$ of the compactified 11d
SUGRA solution to accurately give the tension IIA unstable branes for
all string coupling (only very large coupling).

\section{Unstable branes and $Z_2$ gauge symmetry}

The worldvolume gauge theory of unstable Dp branes (p odd for IIA or
even for IIB) is $U(1)_c\times Z_2$, where $U(1)_c$ is the ``center of
mass'' $U(1)$ and the $Z_2$ acts on the real tachyon (which is
$U(1)_c$ neutral) as $T\rightarrow -T$.  An argument given in
\cite{HKL} for why this $Z_2$ is a discrete gauge rather than global
symmetry follows {}from the construction of type IIA (IIB) unstable Dp
branes by a $(-1)^{F_L}$ orbifold of the Dp-$\overline{\rm Dp}$ brane
system of IIB (IIA): $U(1)_c\times Z_2$ is the subgroup of the
$U(1)_c\times U(1)_r$ gauge symmetry of the Dp-$\overline{\rm Dp}$
brane system which survives under $(-1)^{F_L}$.  Because it's a gauge
transformation, $T\rightarrow -T$ is not really a symmetry but, rather,
a redundancy of these variables.

We can define covariant derivatives $DT$, which transform as $DT
\rightarrow \epsilon (x)T$ under $T(x)\rightarrow \epsilon (x) T(x)$,
with $\epsilon (x)=\pm 1$; $DT$ differs from $dT$ only by terms of
delta function support, needed to cancel the $(d\epsilon) T$ term.
Wilson loops are $Z_2$ group elements, which can only be non-trivial
on a space with nontrivial $\pi _1$ (either via $S^1$ compactification
or by circling a string defect).  Finite energy requires
$DT\rightarrow 0$ at infinity.  As in the vortex case eqn.(1), we
might expect that this correlates the kink number \be {1\over
2T_0}(T(x_t=\infty)-T(x_t=-\infty))={1\over 2T_0}\int dx_t dT \ee with
$Z_2$ flux localized on the codimension 1 domain wall. Then the
absence or presence of $Z_2$ flux is what decides whether the unstable
brane decays to vacuum or a codimension 1 stable brane.

\subsection{Is the $Z_2$ Symmetry Broken?}

We note that the $dT\wedge C$ term (eqn. 2) in the worldvolume of an
unstable Dp brane would break the $Z_2$ gauge invariance unless both
$T$ {\it and} $C$ transform, $T\rightarrow -T$ and $C\rightarrow -C$,
under a gauge transformation.  The fact that
the background field $C$ is charged means that any non-zero $C$
expectation value spontaneously breaks the $Z_2$ gauge symmetry.

The tachyon kink and anti-kink are related by
$T\rightarrow -T$, and thus apparently $Z_2$ gauge equivalent.  Since
the kink and anti-kink correspond to a D($p-1$) brane or anti-brane,
they should not be equivalent.  This can be resolved by saying that
the $Z_2$ gauge symmetry is broken by two effects: the $T$ expectation
value in the kink configuration and also by the fact that
$C$ is charged.  Then the D($p-1$) brane and anti-brane can be inequivalent
and correspond to an alignment or misalignment of these two $Z_2$
breakings.

To get another perspective on the $Z_2$ breaking, consider a
D-$(p+1)$-brane/anti-brane pair with a unit of magnetic flux in the
relative $U(1)$, i.e. a unit of D($p-1$) brane charge.  This system can
decay to an unstable D-$p$-brane, also with one unit of
D-$(p-1)$-brane charge.  Thus the unstable brane must have a kink, and
the kink number cannot change.  This means that, if there is non-zero
D($p-1$) brane charge, the tachyon must be condensed at $\pm \infty$.
Because $T$ is $Z_2$ charged and necessarily condensed, the gauge
symmetry is necessarily broken.  It is not surprising that tachyon
condensation breaks the $Z_2$ gauge symmetry.  But it is surprising
that the unstable brane with non-zero RR charge cannot ever have an
unbroken $Z_2$ gauge symmetry, because the tachyon can never be
uncondensed.

The condensation of the tachyon living on an unstable brane is the
world-volume signature that the unstable brane has decayed to the
vacuum \cite{Senrev}.  The fact that the tachyon must be condensed at
infinity then corresponds to the statement that one cannot meaningfully
discuss an infinite unstable D-$p$-brane.  If it has non-zero
D-$(p-1)$-brane charge, then the tachyon kink must always be non-zero,
which implies that the brane has already decayed to the vacuum, at
least at $\pm \infty$.  From the spacetime point of view this is
because, over an arbitrarily small time $\delta t$, the probability
that the unstable brane decays within an area $A$ can be made
arbitrarily close to 1 simply by increasing $A$.  By the time the
unstable D2-brane has formed, it has already decayed to the vacuum in
an infinite number of places.  For an unstable D-brane, the decay
constant is expected to be on the order of the string scale, i.e. for
$\delta t$ of order the string scale, the probability that the brane
has begun to decay within a string scale volume is $\sim 1/e$.

\subsection{Multiple D-branes From a Single Unstable Brane?}

If a potential has only two minima, then a kink cannot be followed by
another kink; it can only be followed by an anti-kink.  This suggests
that an unstable Dp brane can decay to at most a single D($p-1$)
brane; to get multiple D($p-1$) branes would require multiple unstable
Dp branes (\cite{Wittenperiodic}, \cite{Horavaperiodic}).

On the other hand, a single D($p+1$) brane/anti-brane pair can decay
to an arbitrary integer $n$ D($p-1$) branes when there are $n$ units
of magnetic flux, and thus $n$ vortices.  We can construct domain
walls with any number of flux units, such that the tachyon field
interpolates through zero at the wall and the flux is contained within
the wall; this implies that an unstable Dp brane can decay
to any number of D($p-1$) branes.  A possible resolution is that this
domain wall is really a set of $n$ coincident unstable Dp branes, each
of which decays to a single D($p-1$) brane via a kink.  But note that
a single unstable Dp brane can carry arbitrary fundamental string
charge, which just corresponds to the electric flux of the $U(1)$.
For the case of the unstable D3-brane, these fundamental strings are
wrapped M2-branes, while the D($p-1$)-branes are unwrapped M2-branes.
So we'd then have to explain why an unstable D3-brane can carry
arbitrary wrapped M2-brane charge $n$, but can only have unwrapped
M2-brane charges of magnitude 1 or 0.  This difference can not be
explained by claiming that the $n$ fundamental strings are merely
a single M2 brane wound $n$ times around the dilaton circle: the 
$n$ fundamental strings can be separated with no energy cost, unlike
a multiply wound M2 brane. 

A possible resolution of this puzzle is to conjecture
that the tachyon potential is actually periodic, with an infinite
number of global minima (as in the sine-Gordon model).  In that case,
any number of kinks could appear successively, allowing the unstable
brane to have any integer number of RR charge.  This conjecture seems
to conflict with recent results in boundary string field theory
\cite{SSFT}, which found a non-periodic tachyon potential.  But perhaps 
the domain of the BSFT tachyon actually corresponds to only a finite
segment of a full periodic tachyon potential.

\vskip .25in
{\bf Acknowlegements}

We are grateful to O. Bergman, J. McGreevy, J. Preskill, A. Sen,
S. Shenker, and L. Susskind for useful discussions.  The work of
K. I. and J. K. is supported by DOE-FG03-97ER40546.  M.K. is supported
by Stanford University.

\end{document}